\documentclass[preprint]{aastex}

\shortauthors{Burgasser et al.}
\shorttitle{Gl 570D}

\begin{document}

\title{Discovery of a Brown Dwarf Companion to Gliese 570ABC: A 2MASS T Dwarf
Significantly Cooler than Gliese 229B}

\author{Adam J. Burgasser\altaffilmark{1},
J. Davy Kirkpatrick\altaffilmark{2},
Roc M. Cutri\altaffilmark{2},
Howard McCallon\altaffilmark{2},
Gene Kopan\altaffilmark{2},
John E. Gizis\altaffilmark{3},
James Liebert\altaffilmark{4},
I. Neill Reid\altaffilmark{5},
Michael E. Brown\altaffilmark{6,7},
David G. Monet\altaffilmark{8},
Conard C. Dahn\altaffilmark{8},
Charles A. Beichman\altaffilmark{9},
and Michael F. Skrutskie\altaffilmark{3}}
 
\altaffiltext{1}{Division of Physics, M/S 103-33, 
California Institute of Technology, Pasadena, CA 91125; diver@its.caltech.edu}
\altaffiltext{2}{Infrared Processing and Analysis Center, M/S 100-22, 
California Institute of Technology, Pasadena, CA 91125}
\altaffiltext{3}{Five College Astronomy Department, Department of Physics
and Astronomy, University of Massachusetts, Amherst, MA 01003}
\altaffiltext{4}{Steward Observatory, University of Arizona,
Tucson, AZ 85721}
\altaffiltext{5}{Department of Physics and Astronomy, 
University of Pennsylvania, 209 South 33rd Street, Philadelphia, PA 19104-6396}
\altaffiltext{6}{Division of Geological and Planetary Sciences, M/S 105-21, 
California Institute of Technology, Pasadena, California 91125}
\altaffiltext{7}{Alfred P.\ Sloan Research Fellow}
\altaffiltext{8}{U.S. Naval Observatory, P.O. Box 1149, 
Flagstaff, AZ 86002}
\altaffiltext{9}{Jet Propulsion Laboratory, M/S 180-703, 
4800 Oak Grove Dr., Pasadena, CA, 91109}

\begin{abstract}
We report the discovery of a widely separated (258$\farcs$3$\pm$0$\farcs$4)
T dwarf companion 
to the Gl 570ABC system.
This new component, Gl 570D, was initially identified from
the Two Micron 
All Sky Survey (2MASS).  Its near-infrared spectrum shows the 1.6 and 2.2
$\micron$ CH$_4$ absorption bands characteristic of T dwarfs, 
while its common proper motion with the
Gl 570ABC system confirms companionship.  Gl 570D (M$_J$ = 16.47$\pm$0.07) is 
nearly a
full magnitude dimmer than the only other known T dwarf companion, Gl 229B,  
and estimates of L = (2.8$\pm$0.3)x10$^{-6}$ L$_{\sun}$ and
T$_{eff}$ = 750$\pm$50 K make it significantly cooler and less luminous 
than any other known brown dwarf companion.
Using evolutionary models by Burrows et al.\
and an adopted age of 2-10 Gyr, we derive a mass estimate of 50$\pm$20 M$_{Jup}$
for this object.
\end{abstract}


\section{Introduction}

Direct detection techniques, 
like those that discovered the prototype T dwarf Gl 229B \citep{Na95,Op99}, 
have been used
for the last 15 years to search for brown dwarfs around nearby 
stars\footnote{For a review of these companion searches see \citet{Op99}}.  
Despite the large samples involved, only two {\it bona fide} brown 
dwarf companions have been directly detected, 
Gl 229B and the
young L-type brown dwarf  G 196-3B \citep{Re98}\footnote{The companion object GD 165B
\citep{Be89} may also be a brown dwarf, although its status is questionable
\citep{Ki99b}.}.
Since most of these searches have been confined to a narrow field of view 
around the  
primary (typically 10-60$\arcsec$), widely separated companions\footnote{We
adopt an observational definition for ``widely separated'' as angular 
separation greater than 100$\arcsec$; see \citet{Fi92}.} 
may be missed.  Indeed, both 
 G 196-3B and  Gl 229B are less than 20$\arcsec$ from their primary. 
Field surveys, such as the Two Micron All Sky Survey
\citep[hereafter 2MASS]{Sk97}, the DEep Near Infrared Survey 
\citep[hereafter DENIS]{Ep97}, 
and the Sloan Digital Sky 
Survey \citep[hereafter SDSS]{Yk99}, overcome this limitation. 
Indeed, 
\citet{Ki00} have recently identified 
two 
L-type brown dwarf companions at wide separation.  

We are currently searching the 2MASS catalogs for field T dwarfs 
\citep{Bg98}, brown dwarfs 
spectrally identified by CH$_4$ absorption bands 
at 1.6 and 2.2 $\micron$ \citep{Ki99a}.  
One of our discoveries, 
 2MASSW J1457150-212148 (hereafter  Gl 570D), has been confirmed as a 
widely separated, common proper motion companion 
to the  Gl 570ABC system. This system is comprised of a K4V primary and a M1.5V-M3V close 
binary \citep{Du88,Mi90,Fo99} at a distance of 
5.91$\pm$0.06 pc \citep{Pe97}. 
In $\S$2 we describe the selection of this object from the 2MASS database,
review subsequent observations, 
and establish its common proper motion with  Gl 570ABC.
In $\S$3 we estimate L and T$_{eff}$
of  Gl 570D based on its
distance and brightness, and make T$_{eff}$ and mass 
estimates using the evolutionary models
of \citet{Bu97}.  

\section{Identification of Gl 570D}

\subsection{Selection and Confirmation of Gl 570D}

 Gl 570D was initially selected as a T dwarf candidate 
from the 2MASS Point Source Catalog.
T dwarf candidates were 
required to have J- and H-band detections with J $<$ 16 (2MASS signal-to-noise 
ratio $\sim$ 10 limit), J-H $<$ 0.3 and H-K$_s$ $<$ 0.3 (limit or detection), 
${\mid}b{\mid} > 15^o$ (to eliminate source confusion in the Galactic plane), and no
optical counterpart in the USNO-A catalog \citep{Mo98} 
within 10$\arcsec$.  
Close optical doubles not identified by USNO-A and proper motion stars were
eliminated by examination 
of Digitized Sky Survey (DSS) images of the SERC-J
and AAO SES \citep{Mr92} surveys.  Our search criteria are
also sensitive to minor planets, due to their intrinsically blue near-infrared 
colors \citep{Ve95,Sy99}, lack of optical counterpart at an earlier epoch, 
and point-like appearance
due to the short 
2MASS exposure time (7.8 seconds).
Follow-up near-infrared imaging to eliminate these objects from our candidate pool
was carried out on the
Cerro Tololo InfraRed IMager (CIRIM)
on the Cerro Tololo
Interamerican Observatory (CTIO)
Ritchey-Chretien 1.5m during 1999 July 23-25 (UT).   Gl 570D was 
one of only 11 candidates detected in these observations (the remaining candidates were
likely asteroids).
Optical images
of the  Gl 570D field from the SERC-J and AAO SES surveys, as well
as 2MASS J- and K$_s$-band images, are shown in Figure 1; the  Gl 570ABC triple
can be seen in the lower left corner.  No optical counterpart is seen at either 
the current or projected (proper motion) positions of  Gl 570D, indicating 
very red optical-infrared colors.  
Table 1 lists 2MASS J, H, and K$_s$ magnitudes (rows
[1]-[3]) and colors (rows [4]-[6]) for  Gl 570D, as well as measurements for
G 196-3B and Gl 229B taken from the literature
\citep{Mt96} and from 2MASS data.  
Note that Gl 570D has blue near-infrared colors, similar
to  Gl 229B.

\placefigure{fig-1}
\placetable{tbl-1}

\subsection{Spectral Data}

The 1.6 and 2.2 $\micron$ fundamental 
overtone CH$_4$ bands 
were identified in  Gl 570D from near-infrared spectral 
data taken with the
Ohio State InfraRed Imager/Spectrometer  
\citep[hereafter OSIRIS]{Dp93} on the CTIO Blanco 4m on 1999 July 27 (UT). 
Using OSIRIS's cross-dispersion mode, we 
obtained continuous 1.2-2.3 
$\micron$ spectra with $\lambda$/$\Delta$$\lambda$ $\approx$ 1200.  
The slit width 
was fixed at 1${\farcs}$2 for all observations.  
The object was placed on the slit by direct image centroiding, 
and then stepped across the slit in seven 
positions at 3$\arcsec$ intervals (to offset fringing and detector artifacts) with 
120-second integrations at each position.  A total of 3360 seconds of integration
time was acquired.  
Spectra were then extracted using standard IRAF
reduction packages.  Raw data were 
flat-fielded using observations of the 4m illuminated dome spot and software
generously supplied by R.\ Blum at CTIO.  Object spectra were extracted using a 
template from the A1V standard star HR 5696 \citep{Ho82}.
Wavelength 
calibration was computed from OH sky lines.  Finally, 
telluric corrections and relative flux calibration were done using 
the extracted standard spectrum.

\placefigure{fig-2}

The near-infrared spectrum of  Gl 570D is shown in Figure 2, along with data
for the SDSS T dwarf SDSSp J162414.37+002915.6  
\citep[hereafter SDSS1624+00]{Ss99} obtained on the same night.  Both spectra are
normalized at 1.55 $\micron$, with SDSS1624+00 offset vertically by a constant.   Gl 229B
spectral data from \citet{Ge96}, also normalized at 1.55
$\micron$, are overlain on both 
for comparison (dark line).  
The 1.6 and 2.2 $\micron$ CH$_4$ bands are 
present in all three brown dwarfs,
as well as combined H$_2$O and CH$_4$ absorption from 1.3 to 1.5 $\micron$.
Suppression of flux around 2.1 $\micron$ is likely due to increased H$_2$ absorption
in the low temperature atmospheres \citep{Lz91}.

There is a striking similarity in the spectral morphology of these 
objects; 
however, the
overlaid spectrum of  Gl 229B may indicate some subtle differences. 
There 
appears to be a slight enhancement in flux (relative to  Gl 229B) 
in SDSS1624+00 at
the blue edge of the 1.3 $\micron$ absorption feature and 
at the base of the 1.6 $\micron$
CH$_4$ absorption band.  
Conversely, the spectrum of  Gl 570D does not show these features
and in fact appears slightly
deficient at the 1.5 $\micron$ H$_2$O-CH$_4$ wing and the 2.1
$\micron$ flux peak.  
We might expect such variations if
SDSS1624+00 is warmer than  Gl 229B and  Gl 570D cooler,
as CH$_4$ bands at 1.4 and 1.6 $\micron$ should deepen 
as the observed temperature decreases, since the conversion of CO to CH$_4$ 
will occur at greater optical depth \citep{Bu99}.  Similarly, there should 
be increased H$_2$ absorption in the K-band toward lower temperatures
\citep{Bg99}.  
While metallicity and mixing effects may complicate these simple arguments,
the warmer temperature of SDSS 1624+00 is supported by recent detections of
FeH and CrH bands in its optical spectrum \citep{Li00}
which
are disappearing in the latest L dwarfs \citep{Ki99a}, as well as shallower
H$_2$O and CH$_4$ bands in the near-infrared as compared to Gl 229B \citep{Na00}.
The coolness 
of Gl 570D, based on its association with Gl 570ABC, is discussed below.  

\subsection{Association with Gl 570ABC}

The proximity of the bright Gl 570ABC triple 
led us to suspect possible association for this 2MASS object.  
Fortunately, the system
has a relatively high proper motion of 2$\farcs$012$\pm$0$\farcs$002 yr$^{-1}$
\citep{Pe97}. 
In addition, multiple sampling and the 
2MASS position reconstruction strategy
results in a higher astrometric accuracy\footnote{Cutri, R.\ M., et al. Explanatory 
Supplement to the 2MASS Spring 1999 Incremental Data Release:  
http://www.ipac.caltech.edu/2mass/releases/spr99/doc/explsup.html.}
($\sim$ 0$\farcs$3)
than the raw pixel scale of the 2MASS detectors (2$\arcsec$), sufficient 
to measure the motion of this system on a one-year timescale.  
The original 2MASS scan of the  Gl 570D field was taken on 1998 May 16 (UT);
a second scan was obtained on 1999 July 29 (UT).  
Table 2 summarizes the
resulting astrometric data, indicating that all components have a common sky motion
of 2$\farcs$3$\pm$0$\farcs$4 at position angle 155$\pm$8$\degr$.
The mean motion of all other correlated sources in the same 2MASS scan as
Gl 570D with J $<$ 15.8 ($\approx$ 2000 sources) 
is 0$\farcs$0$\pm$0$\farcs$2 in right ascension and 
0$\farcs$2$\pm$0$\farcs$1 in declination.  This statistically significant common
proper motion confirms companionship.   Gl 570D lies 
258$\farcs$3$\pm$0$\farcs$4 from the K4V primary, a projected physical separation
of 1525$\pm$15 AU.  Note that this is an order of magnitude larger than the A-BC
separation (24$\farcs$7$\pm$0$\farcs$4) and over four orders of magnitude larger
than the B-C separation of 0$\farcs$1507$\pm$0$\farcs$0007
\citep{Fo99}.  
The separation of  Gl 570D is compared to those of 
 G 196-3B and
 Gl 229B in Table 1 (rows [7]-[8]).

The dynamic stability of this system can be 
addressed using the results of
\citet{Eg95} with the separations\footnote{We assume face-on projection and
negligible eccentricity in this order-of-magnitude analysis.} 
listed in Table 2 and masses of 0.7 
M$_{\sun}$ for Gl 570A 
(estimated from the measured mass of the M0Ve eclipsing binary YY Gem; Bopp 1974), 
1.0 M$_{\sun}$ for combined Gl 570BC (directly measured by Forveille et al.\ 1999), 
and 0.05 M$_{\sun}$ for Gl 570D (estimated, as discussed below).  
We find that the system is dynamically stable
for eccentricities less than about 0.6.  
A more rigorous analysis using 
measured orbital parameters is restricted by the
roughly 40,000-year period of Gl 570D around the Gl 570ABC barycenter.

\placetable{tbl-2}

\section{Estimates of the Physical Properties of Gl 570D}

Distance moduli and
absolute J magnitudes for the three brown dwarf companions
 G 196-3B,  Gl 229B, and  Gl 570D, based on the distances to their respective
primaries, are listed in Table 1 (rows [9]-[10]).  
 Gl 570D is nearly a magnitude fainter than  Gl 229B at all three near-infrared
bands.
If 
we assume a  Gl 229B J-band bolometric correction of 2.19$\pm$0.10
\citep{Le99} and a radius 
of (7.0$\pm$0.5)x10$^9$ cm $\approx$ 1 Jupiter radius \citep{Bu93},
we then derive L = (2.8$\pm$0.3)x10$^{-6}$ L$_{\sun}$ and 
T$_{eff}$ = 750$\pm$50 K, roughly 200 K cooler than  Gl 229B, making 
 Gl 570D the 
least luminous and coolest brown dwarf thus
far detected.  More accurate determinations of the effective temperature and mass
of  Gl 570D can be made using brown dwarf evolutionary models, but only if we can
constrain its age ($\tau$).
The proximity of  Gl 570ABC has permitted detailed studies of
kinematic properties, activity, and high energy emission (UV and X-ray),
leading to various age estimates for the system \citep{Le92,Po93,Fl95}.  There 
is a general consensus among these authors that this system is older than 2 Gyr,
which is supported by the lack of activity in the close BC binary \citep{Rd95}.
The solar-like metallicity of  Gl 570ABC \citep{Fo99} and the system's total 
space motion of 
$\approx$ 60 km s$^{-1}$ \citep{Le92} constrains formation to
the Galactic disk, which establishes a rough upper limit of about 10 Gyr.
Using the evolutionary models of \citet{Bu97} and
adopting log (L/L$_{\sun}$) = -5.56$\pm$0.05 and $\tau$ = 6$\pm$4 Gyr,  
we derive values of T$_{eff}$ = 790$\pm$40 K, and M = 50$\pm$20 M$_{Jup}$\footnote{1
M$_{Jup}$  = 1.9x10$^{33}$ grams = 0.0095 M$_{\sun}$.}
(Table 1, rows [13]-[14]).  The effective temperature is consistent with the
brightness estimate above, and is significantly lower than those of 
G 196-3B and
Gl 229B.  Perhaps most interesting is that, despite having the lowest T$_{eff}$,
Gl 570D 
could possibly be the  
most massive of these three brown dwarfs.  This accentuates the difficulty of 
basing comparisons of brown dwarfs on brightness and/or temperature alone, 
and the importance of age determinations in deriving the physical properties of
cool brown dwarfs.
More accurate estimates of this object's properties 
require spectral modeling and additional broad-band photometry, 
and will be addressed in a future paper.

\acknowledgements
A.\ J.\ B.\ acknowledges Robert Blum and Ron Probst for their guidance at the 
telescope and in the reduction process, 
the capable assistance of CTIO telescope 
operators Mauricio Fernandez and Alberto Z{\'{u}}niga, 
useful discussions with Mark Marley, 
helpful comments from the anonymous referee, 
and Daniel Durand for
dealing with high volumes of image requests on the CADC DSS server.  
We thank the 2MASS staff and scientists for their
efforts in creating a truly incredible astronomical resource.
DSS images were obtained from the Canadian Astronomy Data Centre, 
which is operated by the Herzberg Institute of Astrophysics, 
National Research Council of
Canada. 
A.\ J.\ B., J.\ D.\ K., I.\ N.\ R., and J.\ L.\ acknowledge funding through a 
NASA/JPL grant to 2MASS
Core Project science. 
A.\ J.\ B., J.\ D.\ K., R.\ M.\ C., and C.\ A.\ B.\ acknowledge the support 
of the Jet Propulsion
Laboratory, California Institute of Technology, which is operated under
contract with the National Aeronautics and Space Administration.
This publication makes use of data from
the Two Micron All Sky Survey, which is a joint project of the University
of Massachusetts and the Infrared Processing and Analysis Center, funded
by the National Aeronautics and Space Administration and the National
Science Foundation.

\clearpage

\figcaption[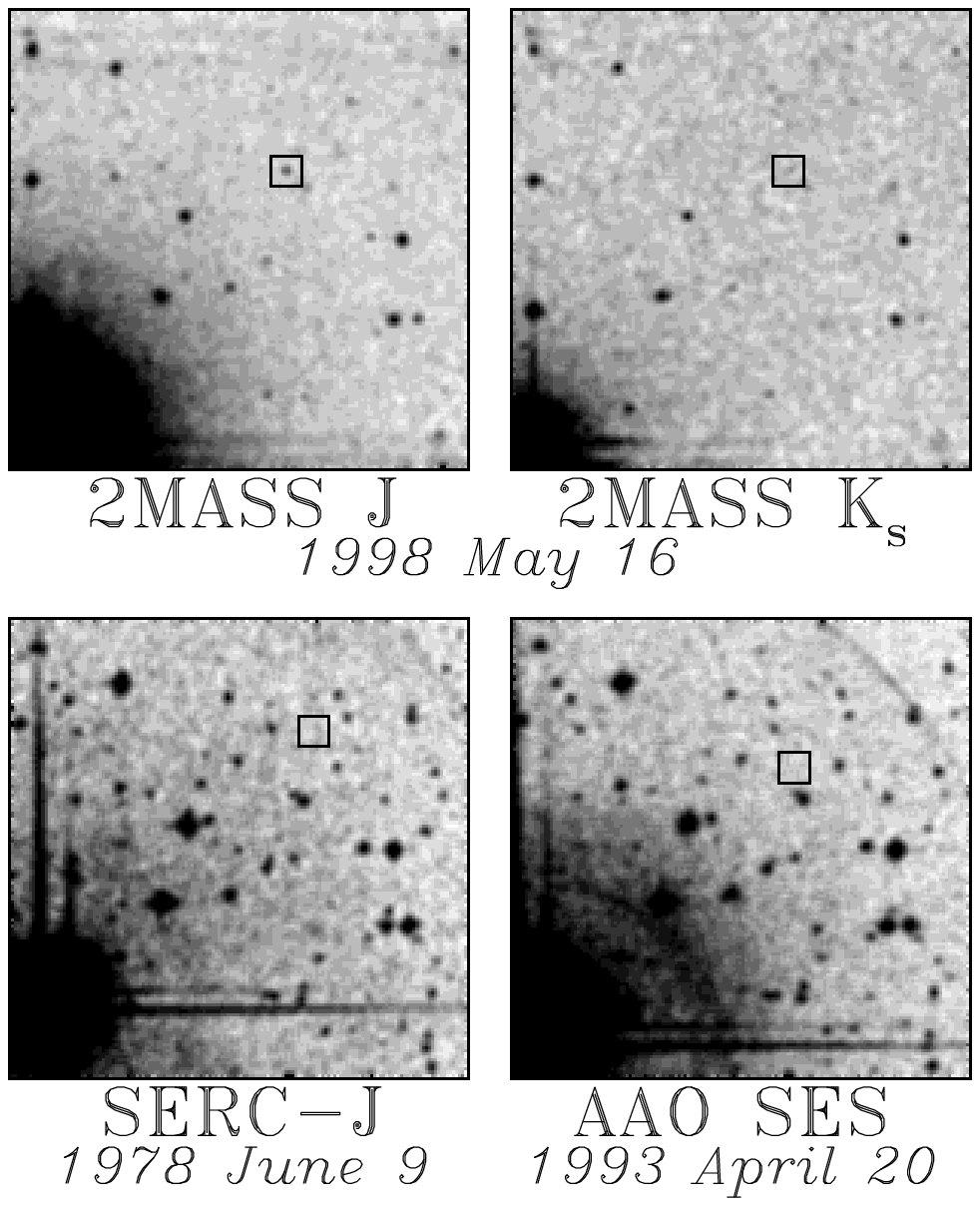]{2MASS J- and K$_s$-band images of Gl 570D, along
with two optical images from SERC-J and AAO-SES at two different epochs.
Each field is 5$\arcmin$ x 5$\arcmin$ with north up and east to the left.
The Gl 570ABC triple is seen in the lower left corner (the BC binary is 
unresolved).
Gl 570D is indicated in the 2MASS images by a 20$\arcsec$ x 20$\arcsec$ box, while its 
projected location due to motion is indicated in each optical image.  No
optical counterpart is seen in either of these images, limiting R-J $\gtrsim$ 6.
\label{fig-1}}

\figcaption[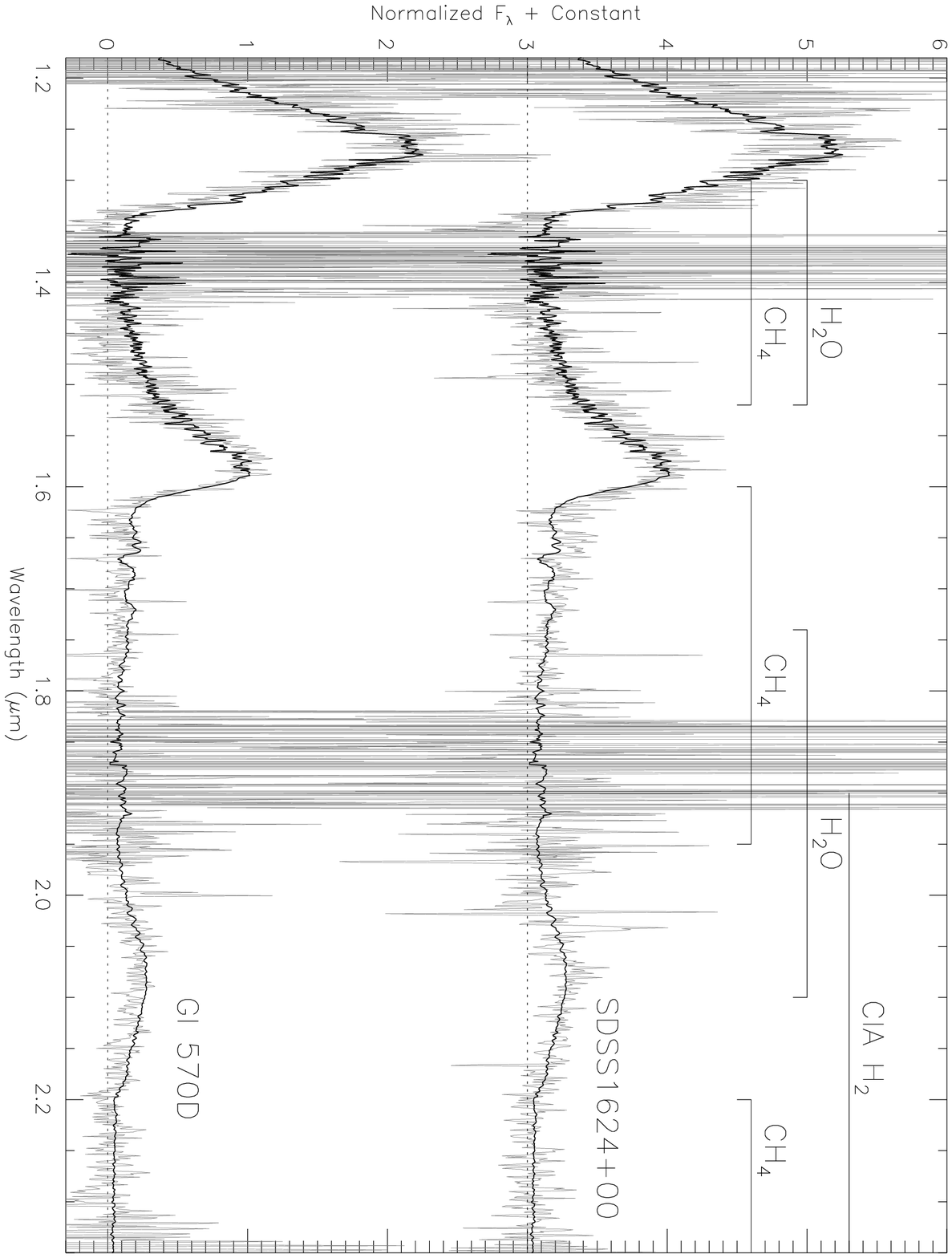]{Near-infrared spectral data for Gl 570D (bottom) and SDSS1624+00
(top).  Both are normalized at 1.55 $\micron$ with spectral data for
Gl 229B from Geballe et al.\ (1996) overlaid 
for comparison (dark line).  An integral offset (3.0) has been
added to the spectra in the top panel to separate them
vertically from the others, and zero levels are indicated
by dashed lines.  The data are strikingly similar,
as CH$_4$ absorption features at 1.3, 1.6, and 2.2 $\micron$ are 
clearly seen in all three objects,
as are broadened H$_2$O absorption bands at 1.3 and 1.9 $\micron$.  
H$_2$ collision-induced absorption (CIA) 
longward of about 1.9 $\micron$ is probably also present.  Despite
the strong similarities, there are some variations in flux near the 
1.3 and 1.5 $\micron$
H$_2$O-CH$_4$ absorption wings, the 1.6 $\micron$ CH$_4$ absorption 
band, and the
2.1 $\micron$ flux peak, all of which could be attributable to 
temperature differences 
between these objects.
\label{fig-2}}

\clearpage

\begin{deluxetable}{clccc}
\tabletypesize{\scriptsize}
\tablenum{1}
\tablewidth{0pt}
\tablecaption{Properties of Confirmed Companion Brown Dwarfs. \label{tbl-1}}

\tablehead{
\colhead{} &
\colhead{Property} &
\colhead{G 196-3B\tablenotemark{a}} &
\colhead{Gl 229B\tablenotemark{b}} &
\colhead{Gl 570D\tablenotemark{c}} 
}
\startdata
(1) & J & 14.90$\pm$0.05 & 14.33$\pm$0.05 & 15.33$\pm$0.05 \\
(2) & H & 13.67$\pm$0.07 & 14.35$\pm$0.05 & 15.28$\pm$0.09 \\
(3) & K$_s$ & 12.81$\pm$0.13 & 14.42$\pm$0.05 & 15.27$\pm$0.17 \\
(4) & J-H & {\phs}1.23$\pm$0.09 & $-$0.02$\pm$0.07 & {\phs}0.05$\pm$0.11 \\
(5) & H-K$_s$ & {\phs}0.86$\pm$0.15 & $-$0.07$\pm$0.07 & {\phs}0.01$\pm$0.11 \\
(6) & J-K$_s$ & {\phs}2.09$\pm$0.14 & $-$0.09$\pm$0.07 & {\phs}0.06$\pm$0.23 \\
(7) & $\rho$, PA ($\arcsec$, $^o$) & 16.2, 210 & 7.8$\pm$0.1, 163 & {\phs}{\phs}258.3$\pm$0.4, 316 (A-D){\phs} \\
  &  & & & {\phs}{\phs}234.1$\pm$0.4, 317 (BC-D) \\
(8) & $\rho$ (AU) & 340$\pm$100 & 44.9$\pm$0.6 & 1525$\pm$15 (A-D){\phs} \\
  &  & & & 1385$\pm$15 (BC-D) \\
(9) & Distance Modulus\tablenotemark{d} & 1.6$\pm$0.6 & $-$1.19$\pm$0.07 & $-$1.14$\pm$0.05 \\
(10) & M$_J$ & 13.3$\pm$0.6 & 15.52$\pm$0.06 & 16.47$\pm$0.07 \\
(11) & log (L/L$_{\sun}$) & $-$3.8$^{+0.2}_{-0.3}$ & $-$5.18$\pm$0.04 & $-$5.56$\pm$0.05\tablenotemark{e} \\
(12) & Age (Gyr) & 0.02-0.3 & 0.5-1.0 & 2-10 \\
(13) & T$_{eff}$ (K) & 1800$\pm$200 & 960$\pm$70 & 750$\pm$50\tablenotemark{f} \\
(14) & M (M$_{Jup}$) & 25$^{+15}_{-10}$ & 43$\pm$12 & 50$\pm$20\tablenotemark{f} \\
 
\tablenotetext{a}{Photometry from 2MASS; remaining data from \citet{Re98}.}
\tablenotetext{b}{Data from \citet{Na95,Ma96,Le99}.}
\tablenotetext{c}{Data for epoch 1998 May 16 (UT).}
\tablenotetext{d}{Data for Gl 229A and Gl 570A from Hipparcos \citep{Pe97}.}
\tablenotetext{e}{Assuming J-band bolometric correction of 2.19$\pm$0.10 from
\citet{Le99}.}
\tablenotetext{f}{Derived from evolutionary models by \citet{Bu97}.}

\enddata
\end{deluxetable}

\clearpage

\begin{deluxetable}{clcccccc}
\tabletypesize{\scriptsize}
\tablenum{2}
\tablewidth{0pt}
\tablecaption{2MASS Astrometry for Gl 570ABCD. \label{tbl-2}}

\tablehead{
 & 
\multicolumn{2}{c}{1998 May 16 (UT)} & 
\multicolumn{2}{c}{1999 July 29 (UT)} & 
\multicolumn{2}{c}{Difference} & \\
\colhead{Component} &
\colhead{RA\tablenotemark{a}} & 
\colhead{Decl.} &
\colhead{RA} &
\colhead{Decl.} &
\colhead{$\arcsec$} &
\colhead{$^o$} 
}

\startdata
A & 14:57:27.87 & -21:24:52.72 & 14:57:27.93 & -21:24:54.87 & 2.3$\pm$0.4 & 159$\pm$8 \\
BC & 14:57:26.42 & -21:24:38.54 & 14:57:26.49 & -21:24:40.77 & 2.4$\pm$0.4 & 156$\pm$7 \\
D & 14:57:14.96 & -21:21:47.79 & 14:57:15.04 & -21:21:49.82 & 2.3$\pm$0.4 & 151$\pm$8 \\
\tablenotetext{a}{All coordinates are equinox J2000.0.}
\enddata
\end{deluxetable}

\clearpage

\plotone{fig1.ps}

\plotone{fig2.ps}

\end{document}